# Modeling overall energy consumption in Wireless Sensor Networks

Najmeh Kamyabpour
Advanced Research in Networking Laboratory
iNEXT-Centre for Innovation in IT Services and
Applications, University of Technology, Sydney Broadway
NSW 2007 Australia
najmeh@it.uts.edu.au

Doan B.Hoang
Advanced Research in Networking Laboratory
iNEXT-Centre for Innovation in IT Services and
Applications, University of Technology, Sydney Broadway
NSW 2007 Australia
dhoang@it.uts.edu.au

*Abstract-* **Minimizing the energy consumption of a wireless sensor network application is crucial for effective realization of the intended application in terms of cost, lifetime, and functionality. However, the minimizing task is hardly possible as no overall energy cost function is available for optimization. Optimizing a specific component of the total energy cost does not help in reducing the total energy cost as this reduction may be negated by an increase in the energy consumption of other components of the application. Recently we proposed Hierarchy Energy Driven Architecture as a robust architecture that takes into account all principal energy constituents of wireless sensor network applications. Based on the proposed architecture, this paper presents a single overall model and proposes a feasible formulation to express the overall energy consumption of a generic wireless sensor network application in terms of its energy constituents. The formulation offers a concrete expression for evaluating the performance of a wireless sensor network application, optimizing its constituent's operations, and designing more energy-efficient applications. The paper also presents simulation results to demonstrate the feasibility of our model and energy formulation.**

*Keywords: Sensor, Wireless Sensor Networks (WSNs), Hierarchy Energy Driven Architecture (HEDA), Overall Energy Consumption Model.*

## I. INTRODUCTION

Energy consumption is easily one of the most fundamental but crucial factor determining the success of the deployment of sensors and wireless sensor networks (WSNs) due to many severe constraints such as the size of sensors, the unavailability of a power source, and inaccessibility of the location and hence no further handling of sensor devices once they are deployed. Efforts have been made to minimize the energy consumption of wireless sensor networks and lengthen their useful lifetime at different levels and approaches. Some approaches aim to minimize the energy consumption of sensor itself at its operating level [1] , some aim at minimizing the energy spent in the input/output operations at data transmission levels,[2] ; and others target the formulation of sensor networks in terms of their topology and related routing mechanisms [3]. The generic goal here is to reduce the amount of energy consumption of some components of the application as much as possible by reducing the tasks that have to be performed by the sensors and the associated networks yet still fulfill the goal of the intended application. In addition to the minimization effort, some approaches tried to replenish the energy capacity of the sensors by building into them components and mechanisms for harvesting additional energy from available energy sources while operating within their environments such as solar, thermal, or wind power sources, [4]. Yet another approach is to scan systematically through the levels OSI network reference model and minimize the energy consumption at some level (if feasible) with the hope that this will reduce the overall energy consumption of the entire network and the application [5].The main problem with these approaches is that they may succeed in reducing the energy consumption in one component of the overall WSN application, but this gain is often negated by an increase in the energy consumed in other components of the application. There has been very little understanding of overall energy consumption map of the entire application, the major components of this energy map and the interplay among the components.

We have approached the problem for a different angle by focusing of energy constituents of an entire sensor network application. An energy constituent represents a major energy-consuming entity that may be attributed to a group of functional tasks. Eventually, these tasks have to be mapped to energy consumed actions that have to be performed by sensors and other components such as sensors' antennas, transceivers and central processing units.

The ultimate aim is to produce an energy map architecture of a generic WSN application with essential and definable energy constituents and the relationship among these constituents so that one can explore strategies for minimizing the overall energy consumption of the entire application. The HEDA architecture as proposed recently is the result of our effort in this direction. Based on this architecture, this paper proposes a formulation of the energy consumption of an entire application in terms of mathematical expressions that enable one to analyze and optimize the energy consumption function. Our architecture focuses on energy constituents rather than network layers or physical components. Importantly, it allows the identification and mapping of energy-consuming entities in a WSN application to energy-constituents of the architecture. Specifically, in this paper we do not only identify constituents a WSN application but we also identify individual components and their contribution to each of the constituents of HEDA architecture. Energy consumption of the constituent is formulated in terms of its components. Furthermore, we identify and take into account in the mathematical expressions salient parameters (or factors) that are believed to play a significant role in an energy component. Preliminary simulation results are also presented to demonstrate the feasibility of the model for further study and evaluation.

## II. OVERALL ENERGY CONSUMPTION FORMULATION

In [6] we proposed a new model where energy is the focus. The model is called HEDA. In this model, we identify major energy consumption components in terms of their roles/activities relative to the network and with respect to the application and model the whole wireless sensor network accordingly. This section summarizes HEDA architecture and its features. Based on the proposed architecture, we formulate the overall energy consumption of the entire set up based on HEDA architecture. In particular, we will express the energy consumption of each HEDA's constituent by its principal parameters or components. The overall energy consumption of the entire systems is expressed in terms of relationship among constituents.

We suppose a continuous time between t1 and t2 for the energy consumption measurement. Residual energy in time t is defined by omitting consumed energy in Δt from the initial battery power in t-Δt. Thus, the energy consumption will be determined in Δt.

$$E_{residual,i}(t) = E_{initial,i}(t - \Delta t) - E_{consumed,i}(\Delta t) \quad (1)$$

$$E(\Delta t) = \frac{\partial E}{\partial t} \Delta t$$

$$\Delta t = t_2 - t_1$$

Realistically, we anticipate a nonlinear relationship between the overall energy consumption of the system and its constituents depending on the application and the overall design. However, this nonlinear formulation requires more extensive exploration as we do not understand enough the metric associated with the energy of each constituent and we are unsure about the mathematical models that can handle such a non-linear relationship. For this paper, a simpler linear approach is adopted to model the overall energy consumption and explore the implication. Future work will explore nonlinear approaches. In the following the overall energy is expressed as a linear combination of HEDA's constituents. Interplay among the components can be taken into account in terms of their weights as some function of the design of the WSN and the application.

The total energy consumption of node *i* in the interval Δ*t* based on constituent of HEDA as follows:

$$E_{consumed,i}(\Delta t) = \lambda_1 E_{individual,i}(\Delta t) + \lambda_2 E_{local,i}(\Delta t) + \lambda_3 E_{global,i}(\Delta t) + \lambda_4 E_{battery,i}(\Delta t) + \lambda_5 E_{snk,i}(\Delta t) \quad (2)$$

$subject\ to:$

$1: E_{local,i} > 0$

$2: E_{global,i} > 0$

$3: \lambda_1 E_{individual,i}(\Delta t) + \lambda_2 E_{local,i}(\Delta t) + \lambda_3 E_{global,i}(\Delta t) + \lambda_5 E_{snk,i}(\Delta t) < \lambda_4 E_{battery,i}(\Delta t)$

The first constrain expresses condition for necessity to establish a collaboration connection. The second constrain shows the necessary and sufficient condition for accessibility of the node in the network. The third constrain means a node should have enough energy to do network tasks otherwise it is not active and should be removed from the network calculations. Each constituent is expressed in terms of key parameters (or factors). These key factors are determined based on application requirements. On the other hand, these parameters may influence more than a single constituent; hence energy constituents may partially overlap. Consequently, the interplay among energy constituents must be taken into account in evaluating the overall energy consumption of the entire setup. For example, the number of neighbors determined by topology in the global constituent has direct influence in energy consumption of the local constituent. We will elaborate on the model for each of the constituents in the following sub-sections.

*A. Individual Constituent*

The individual constituent can be a state-based constituent, because every unit has different energy level consumption in different states (figure 1). In addition, this constituent involves two different types of transitions: transitions between units and transition between states of a single unit. The overall energy consumption in individual constituents is expressed as follows:

$$E_{individual,i}(\Delta t) = \sum_{u=1}^{N_u} \sum_{s \in S} \sum_{w \in W} I(e_{u,s}, e_{u,w}, t_{u,s}) \quad (3)$$

$$\sum_{s \in S} e_{u,s} > \sum_{w \in W} e_{u,w} \qquad u \in U$$

Since most of energy minimization methodologies use idle and sleep states for avoid of wasting energy in idle states, the above constraint states that the total energy consumed for switching among states should be smaller than the total energy consumption of states.

Energy consumption in an active state for each unit depends on several factors as follows:

$$e_{1,active}(\Delta t) = F_1(f, b_{proc}) \quad (4)$$

According to Eqn.4, the energy consumption of the processor unit in an active state depends on the number of processed bits and the frequency based on the following equation[7][9]:

$$p \propto cv^2 f \quad (5)$$

This proportionality expresses that the energy consumption of the processor is proportional to the voltage and the frequency of the operation. Since the frequency and the voltage can be related. We consider frequency as an effective parameter in this unit.

$$e_{2,active}(\Delta t) = F_2(r_{sense}, g_{sense}, b_{sense}) \quad (7)$$

Eqn.7 shows that the energy consumption of a sensor unit in an active state depends on the sensor radius, the data generation rate, and the number of generated bits.

$$e_{3,active}(\Delta t) = F_3(b_{store}, e(rd), e(wt), t_{store}) \quad (8)$$

Energy consumption of a memory unit in an active state depends on the number of stored bits, the number of memory read and write, and the duration of storage.

$$e_{4,active}(\Delta t) = F_4(b_{Rx}, b_{Tx}, e(code), e(dcode)) \quad (9)$$

Energy consumption of the transceiver unit for digital signal processing in an active state depends on the number of received and transmitted bits, and the amount of needed energy for coding and decoding packets.

The energy wastage in idle and sleep states can be measured according to the base amount of energy consumption in these states which depends on unit type and duration of staying in the state[6]. More over switching among the unit's states also consumes considerable amount of energy, this energy is measured differently for different type of unit.

Task transitions between units shown in figure 1 affect the level of energy consumption of the individual constituent. Explicitly, figure 2 shows an example of a data generation sequence in the individual constituent. Data generation time (sensing time), process time, storage time, and data transmission time may all contribute to the overall energy consumption of the individual constituent. These parameters can determine the number of task transmissions between units. For example if the data generation time is smaller than the process time, the number of memory read and write will increase because the data should be stored until processor completes the task. Also if the process time is smaller than the transmissions delay then memory read and write will increase. Limited resources of a sensor such as memory units should be used carefully. For instance if sensor does not have enough memory it can not process received packets. So we need to optimize parameters of each unit with respect to the parameters of other units. Therefore the active time in each constituent is one of the important factors in energy consumption of other units.

### B. Group Energy Consumption

Generally, transmission is a key task in communication among nodes. Energy consumption for packet transporting in the network is in proportion to the distance. The distance to neighbors can increase or decrease the energy consumption of radio channel to transmit a data bit. Heinzelman et al. [8] derived the energy consumption of transmit and receive a k-bit message for a microsensor. The needed energy for the transmit amplifier to send a bit is shown as $e_{amp}$. Hence, in local and global constituents, the energy consumption for transmitting $k$ bits to a node of distance $d$ from the transmitting node is defined as follows:

$$E_{Tx}(d) = e_{amp} d^2 k \qquad (10)$$

and energy consumption of receiving k bits from a node is proportional to the receiver electronics energy per bit, $e_{elec}$ is defined as [8]: $\qquad E_{Rx} = e_{elec} k \qquad (11)$

These equations are general forms of the energy consumption for communication. The important factors, which increase or decrease the energy consumption of transmission and receive operations, should be considered by network designers. Determining the number and the distance to neighbors, the transmission rate, the receive rate, the optimum size of data and message packets are all important in determining the amount of energy consumption in the radio channel. Each factor is thus considered in a component of a constituent of HEDA architecture. Although the transmit amplifier is shared among Group constituents, its energy consumption is determined based on its different roles in different constituents. The following is the discussion of each constituent:

*Local Constituent:* The local communication is concerned with initiating and maintaining all communications between a sensor node and its immediate neighbors so that they can co-exist to perform the roles within the WSN as dictated by the objective of the application. The following equation shows local energy consumption of a node in interval time $\Delta t$:

$$E_{local,i}(\Delta t) = \sum_{j \in neighbour_i} L \begin{pmatrix} e_{ij}(mon), e_{ij}(\sec), e_i(idle), \\ e_{ij}(local), e_{ij}(coll), e_i(ohear) \end{pmatrix} \qquad (12)$$

$1: neighbour_i \geq 1$

$2: e_{ij}(local) < e_i(idle) + e_{ij}(coll) + e_i(ohear)$

The first constrain shows that the node has to have at least one neighbor to be able to relay data and exist in the network. The second constrain is the condition for having optimum energy consumption in the local. This means that energy consumed for control packets of the local protocol which aim to manage effectively access to the shared media in order to avoid collision, idle listening and overhearing should not be bigger than sum of energy consumption of these costly problems in the network when there is no management on the shared media.

$$e_{ij}(mon) = F_5(d_{ij}, b_{mon}, r_{Tx}) \qquad (13)$$

Neighbour monitoring is used for gathering information of neighbour's available resources such as residual energy and memory space. Therefore energy consumption depends on the distance of neighbours and the number of exchange bits. $d_{ij}$ is distance between node $i$ and its neighbour $j$ and $b_{mon}$ is the number of exchange bits between them, $r_{Tx}$ is the transmission radius and the number of neighbours is proportional to $r_{Tx}$.

$$e_{ij}(\sec) = F_6(d_{ij}, b_{\sec}, r_{Tx}) \qquad (14)$$

Security management is for preventing malicious nodes from destroying the connectivity of the network and tampering with the data. Energy consumption depends on the distance of neighbours and the number of exchange bits. $b_{sec}$ is the number of exchange bits between node $i$ and its neighbour $j$,

$$e_{ij}(local) = F_7(d_{ij}, b_{local}) \qquad (15)$$

Various local communication protocols have to be performed to maintain the node's relationship with its neighbours. This type of protocol overheads must be taken into account in terms of energy consumption. Energy consumption depends on the distance of neighbours and the number of exchange bits. $b_{local}$ is the number of exchange bits between node $i$ and its neighbour $j$.

$$e_{ij}(coll) = F_8(d_{ij}, b_{reTx}, n_i, g_{Tx}, r_{Tx}, net_{dens}) \qquad (16)$$

If the node does not receive an acknowledgment for the transmitted packet, it has to retransmit the packet. This situation happens when neighbours transmit packets on the shared medium at the same time. In this case, some parameters come in to consideration: the distance of neighbour, the number of retransmitted bits, the number of neighbors, and the data transmission rate. . $d_{ij}$ is distance between node $i$ and its neighbour $j$ and $b_{reTx}$ is the number of retransmitted bits between them. $n_i$ is the number of neighbour of node $i$ and $g_{Tx}$ is the transmission rate of node $i$, $r_{Tx}$ is the transmission radius. The network density, $net_{dens}$, may increase or decrease the probability of collision.

$$e_i(ohear) = F_9(b_{ohear}, r_{Tx}, net_{dens}) \qquad (17)$$

The node receives packets that are sent to the shared medium. Even the node is not the destination, it still has to examine the packet to figure out what to do. Energy consumption depends on the number of overheard bits. $b_{ohear}$ determines the number of overheard bits in node i, The network density, $net_{dens}$, may increase or decrease the probability of collision overheard packets.

*Global Constituent:* The global constituent is concerned with the maintenance of the whole network, the selection of a suitable topology and the routing strategy employed. This may include energy wastage from packet retransmissions due to congestion and packet errors. The global constituent is defined as a function of energy consumption for topology management, packet routing, packet loss, and protocol overheads.

$$E_{global,i}(\Delta t) = G\begin{pmatrix} e_i(topo), e_i(rout), \\ e_i(global), e_i(pktls) \end{pmatrix} \quad (18)$$

$1: e_i(rout) > 0$

$2: e_i(rout) > e_i(topo) + e_i(global) + e_i(pktls)$

The first constrain shows that there is at least a path from node $i$ to the destination within the network and the node participates in the global communication. The next constraint shows that the energy consumed for control packets and retransmitted packet should be smaller than the routed data packets from an effective energy consumption point of view otherwise this constituent wastes the node's energy.

$$e_i(topo) = F_{10}(a_i, d_{iA}, b_{topo}, n(t)) \quad (20)$$

Where $e_i(topo)$ represents the energy consumption for establishing a relevant topology through the nodes based on the application's objective. $a_i$ is number of nodes accessible nodes for node $i$, $d_{iA}$ is distance node $i$ and an accessible node, $b_{topo}$ represents the number of exchange bits for topology management.

$$e_i(rout) = F_{11}(n(t), d_{iD}, h_{iD}, b_{rout}) \quad (21)$$

$e_i(rout)$ represents the energy consumption for determining and maintaining hops and transporting packets to the destination. The number of relaying hops can be expressed as a cost component in term of energy dissipation. It should be determined and minimized by a suitable routing method. The cost for maintaining the network connectivity should also be accounted for if hops fail during the network life time. $n(\Delta t)$ determines the number of active nodes in the network in interval time $\Delta t$. This may be useful for selecting the best routing method during the network lifetime. Therefore routing method can be determined dynamically according to the current network situation. $d_{iD}$, $h_{iD}$ are respectively distance between node i and the destination and the number of hops between node i and the destination via neighbors and they may help to select the best path.

$$e_i(global) = F_{12}(d_i, b_{global}) \quad (22)$$

$e_i(global)$ represents the energy consumption due to protocol overheads. It is calculated based on the cost transporting control packets for maintaining the overall network topology and configuration. $d_i$ is the distance between node $i$ and its neighbour and $b_{global}$ is the number of exchange bits between neighbours.

$$e_i(pktls) = F_{13}(d_i, b_{pktls}) \quad (23)$$

$e_i(pktls)$ represents the energy consumption due to packet loss. Selecting inappropriate topologies and routing methods may cause congestion and packet-loss in the network. In this case, extra energy consumption has to be added if a node is required to retransmit a packet. $d_i$ is distance between node $i$ and its neighbour and $b_{pktls}$ is the number of exchange bits between neighbours.

*C. Environment Constituent*

In cases where nodes are capable of extracting or harvesting energy from the environment, we propose to take into account this positive energy component in determining the lifetime of the WSN. The environment constituent as a positive energy component can be formulated as follows:

$$E_{battery,i}(\Delta t) = -H_i(t) \quad (24)$$

Where P(t) is amount of harvested energy at time t.

*D. Sink Constituent*

Energy consumption of nodes from the sink constituent viewpoint can be formulated as follows:

$$E_{snk,i}(\Delta t) = K(e_i(snk)) \quad (25)$$

Where $e(k)$ shows consumed energy of each node to communicate with the sink and perform sink's commands.

$$e_i(snk) = F_{14}(b_{snk}) \quad (26)$$

The above equation means energy consumption of node $i$ for sink constituent depends on number of received bits from the sink.

### III. EXPERIMENTAL RESULTS

In this section the aim is to conduct the range of simulation experiments to evaluate the residual energy in the network with respect to the different constituents of HDEA. Because events in the network occur in intervals in millisecond and initial power of sensors is limit, the vitality of the network is usually one to two minutes. Therefore we evaluated residual energy of wireless sensor application during an interval sixty seconds. In particular we will focus on the individual, the local, and the global constituents. To gain a better understanding of the energy contribution (consumption) of these constituents and their main parameters, we focus at this stage several parameters that are believed to play significant roles in the overall energy consumption. For the individual constituent, we select the sensor's sensing radius as it determines the coverage of the sensor field. For the local constituent, we select the transmission radius of a sensor as it concerns the number of neighbours. For the global constituent, we select the routing scheme as it affects the data transport from sensors to sinks. We will investigate the influence of the individual, the local and the global constituents by measuring the residual energy and the energy consumption of the network. We consider the sensor radius, the sensor transmission radius, and the routing scheme as variable in our simulation experiments while keeping all other parameters fixed. We compared variations of residual energy for different constituent's parameter to obtain the best result for an energy consuming component of a constituent.

We simulated an application based on constituents of HEDA architecture. We assumed 100 sensors for our simulation. They are deployed in a 500*500 pixel area (figure 3). They generate data from environment events which occur at random times and places in the area. We considered effective parameters on energy consumption of process, memory and radio units as constant in Individual constituent of all sensors. For our experiments, sensor radius was considered as an effective parameter of sensor unit; other parameters of the individual constituent such as the sensing rate and the costs for different states of various units are kept constant for selective study. The influence of different sensor radius was measured on overall residual energy of the network. Also, the considered variation of sensor radius parameter is the same for all sensors in the network.

As for the local constituent parameters, number of the energy-consuming bits required for maintaining individual sensor's local environment and the network density were kept constant for the duration of the experiment, but the sensor transmission radius was varied. Neighbor selection usually is application dependent and a node placed in the covered area of another node may be chosen as neighbor of that node. In our application number of neighbors was changed based on the variation of transmission rate. Figure 4 shows how number of neighbors varies for different transmission radius and network density. In addition, to be realistic the cost of distance (Eqn.10) is considered in transporting packets through to the network.

For the global constituent we consider the routing method as the variable of interest. Since topology and routing are costly and significant energy consuming components of the global constituent, they will certainly play main roles in determining the residual energy of the network. Increasing the transmission radius increases of number of possible connections of each node and decreases the number of hops from nodes to their sinks. Nodes can establish connections with all nodes located in area reached by node's transmission radius and the type of connection among nodes was determined based on geographic positions of nodes and sinks. We define two types of connection: sender and receiver. Sender connection of node is a connection that node send data via it. Receiver connection means a connection that the node only receives data from it. Therefore nodes have knowledge of position of their neighbors and position of the sinks. They choose a sender connection with nodes which are nearer to the sinks. In our experiments, two different routing methods were considered: Selective and Random. Selective routing method is based on residual energy of nodes and busy degree of nodes [6]. In random method nodes select randomly a sender connection to send data to the sinks.

In this application nodes did not have the capability of extracting energy from their environment and they had only an initial power. They consume their initial power for surviving in the network. For this reason we did not consider the energy contribution of the environment constituent. Sink may be one or a group of powerful nodes and can be applied in any place in the application's area, we were deployed a group of nodes as sink in special section of the area (the left side of figure 3); because in this case, it is easy to initialize, maintain, manage, and control network connections and paths to the sinks. Sinks did not manage or control sensors in the network; they just have responsibility for collecting the received packets. Therefore, the energy consumption of the sink constituent is not considered.

Figure 5 shows the variation of residual energy for four different sensor radiuses with constant transmission radius and the used routing method is selective. According to the diagram the maximum and minimum residual energy respectively belong to the sensing radius 30 and 60. Because of shared sensing area if we increase the sensing radius, data redundancy in the network increases and consequently energy consumption of routing increases. So that energy consumption of local constituent increase and network energy drops dramatically. As a result, with the possible smallest sensing radius which covers the application environment, optimum energy consumption is obtained with respect selective routing method and constant transmission radius. We repeated this test for the random routing method (figure 6). The result is similar to the selective method test and the smallest sensing radius causes the biggest residual energy. Hence increase of sensing radius causes $e_{sense, active}$ goes up due to increasing the duration of active state of sensing unit and also $e_{local,i}$ rises because of increasing e(rout), thus in the both tests we expected that increasing the sensing parameter decrease the network's residual energy.

In figure 7, we considered the transmission radius as an effective parameter of local constituent. Because the position of nodes is not changed during the tests and nodes should have at list a connection with another node in the network we have to find a base value for the transmission radius. In this network the limit of transmission radius is $r_{Tx}>=130$ in order to have a connected network at initial time. The sensing radius was considered constant ($r_{Rx} = 40$) during the test and the applied routing method is selective. As can be seen, Residual energy has different variation in comparison with figure 5 with the same sensing radius ($r_{sense}= 40$) for different transmission radius. Figure 8 is the same test but by random routing method. In comparing figure 7 ($r_{sense} = 40$) with figure 8, the network has different residual energy with respect the variations of the transmission radius during the test. Generally increasing the transmission radius results rising the number of neighbors ($neighbour_i$) and defines new neighbors in longer distance. Therefore collaboration with these new neighbors is costly. As a consequence, $E_{local,i}$ increases.

On the other hand, increasing of transmission radius creates new paths with smaller number of hops, accordingly it decrease energy dissipation in the network. Thus the cost increases because of increasing distances and on the other hand we face with decreasing number of hop and we do not also know which one have more weight in consuming energy. The behavior of the network for these two routing method shows how these parameters (number of hop, distance) have different influences on residual energy. For tests with selective method decrease of number of hops and related paths keep network connected and the application performs as effective for longer time, in contrast in random routing tests because load of the network is not controlled on short paths and nodes deployed nearer to the sinks and nodes spent more energy because of cost of distance in these the network disconnected from sink area early. Difference of the residual energy of random and selective method is because of energy of inaccessible nodes are counted in overall residual energy. We controlled this situation in our model by considering constrains 1 and 2 in the overall energy consumption (Eqn.2), constrain 1 of the local (Eqn.12) and global (Eqn.18) constituents. But in these tests energy consumption of inactive nodes counted in overall energy consumption because our aim is not to compare these routing methods and the aim is to show how the constituent's parameters affect on overall energy consumption.

Figures 9 and 10 show the change of overall residual energy of the network based on variation of parameters of the individual and local constituents, for random and selective routing methods, respectively. As can be seen from these figures, the biggest value of the residual energy is belonged to smallest transmission and sensing radius for selective method and it is almost equal for smallest transmission radius and different value of sensing radius for random method. The reasons are as the following: increasing the sensing radius results in increasing $e_{sense, active}$ (due to increase of covered sensing area) and e(rout) (due to the increase of data redundancy) and therefore raising $E_{individual,i}$ and $E_{global,i}$, respectively. Moreover, increasing Transmission radius rises $E_{local,i}$ and $E_{global,i}$ indirectly by increasing $neigbour_i$ and $d_{ij}$ and $h_{iD}$, in that order; however, larger Transmission radius results in higher $d_{iA}$ and consequently more $E_{global,i}$. As a general rule, selective routing method spends more global energy than random routing method ($E_{global,i}$ (selective) > $E_{global,i}$ (random)) due to maintenance connectivity, choice based on residual energy and busy degree, so e(rout) as energy consuming component defined in global constituent, of selective method is bigger than zero for longer time than random method. These reasons cause that the residual energy varies based on sensing and transmission radius and routing methods as shown in figure 9 and 10.

$$E_{consumd,i}(\Delta t) = E_{initial,i}(t - \Delta t) - E_{residual,i}(t) \qquad (27)$$

According to the above results overall energy consumption and Eqn.27 resulted from the Eqn.1, $E_{consumed,i}$, is proportional to individual, local, global constituents as parameters of these constituents have direct effect on the overall energy consumption and also their energy consumption. Therefore for manipulating the overall energy consumption, according to the Eqn.1 and achieved results, interactions, overlaps and influences of all constituents should be taken in to account.

## IV. CONCLUSION AND FUTURE WORKS

In this paper, we presented a new approach for minimizing the total energy consumption of wireless sensor network applications based on the Hierarchy Energy Driven Architecture. In particular, we identified components of each constituent of HEDA. We extracted a model for each of the constituents and components in terms of their dominant factors (or parameters). We proposed a formulation for the total energy cost function in terms of their constituents. Simulation results for lifetime and residual energy of a sample network with different sensor radius, transmission radius and random and selective networks demonstrated that our model and formulation can be used to optimize the overall energy consumption, determine the contribution of each constituents and their relative significance. The implication is that optimizing the energy of the general model with respect to all constituent parameters will enable one to engineer a balance of energy dissipation among constituents, optimize the energy consumption among them and sustain the network lifetime for the intended application. It should be noted that many important issues are still to be explored. This paper only suggested an outline model for each constituents; a detailed energy model for each of the constituent of HEDA is to be studied. The paper identified a number of dominant parameters of each energy components, however, not all features of WSNs have been taken into consideration and they should be explored and investigated thoroughly.

Clearly, the relationship among the energy constituents and their interplay within an application are important; we plan to explore the patterns and the shape of the energy consumption for a generic application and produce a comprehensive map of energy consumption relative to a specific application. Preliminary investigation assumed a weighted linear combination of energy consumption of the constituents, in the future, we plan to produce a more accurate energy cost function which accurately place due emphasis on parameters, components and the playoff factors among components. We believe that a non-linear cost function rather than a simple linear combination would allow the model to adapt better to a specific WSN application.

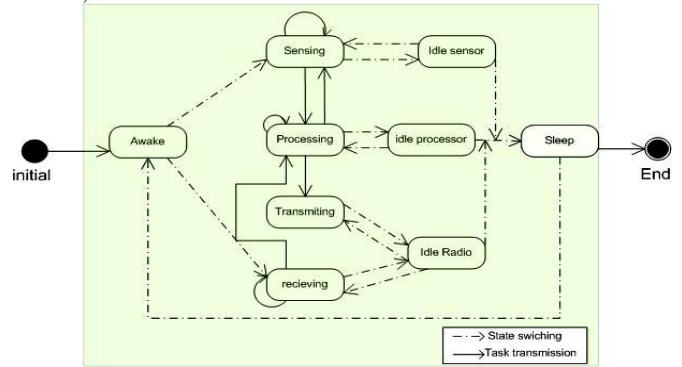

*Figure 1. General State Diagram of the individual constituent.*

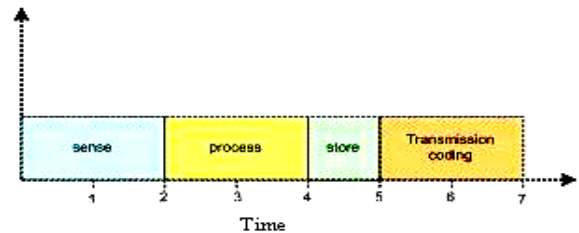

*Figure 2. An example of the data generation sequence in individual constituent*

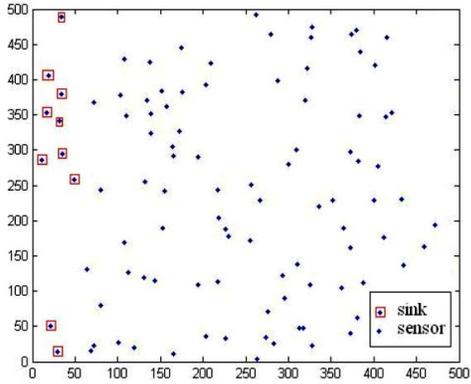

*Figure 3. Randomly deployed sensors and sinks in the application*

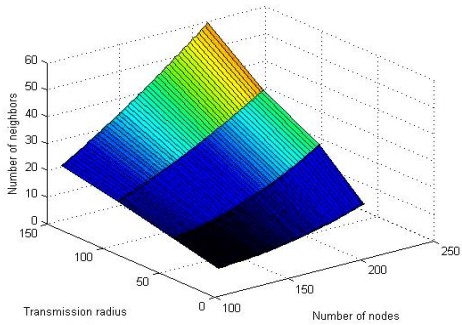

*Figure 4. Maximum number of neighbours for different transmission radius and network size*

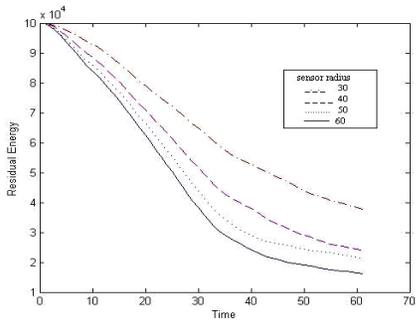

Figure 5. Residual energy for different sensor radius ($r_{Tx}$=130, Selective)

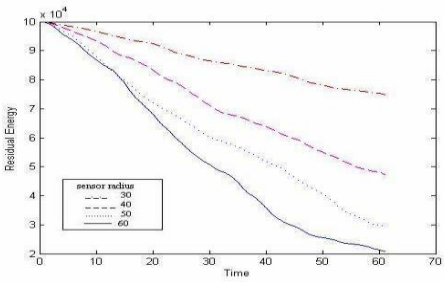

*Figure6. Residual energy for different sensor radius ($r_{Tx}$=130, Random)*

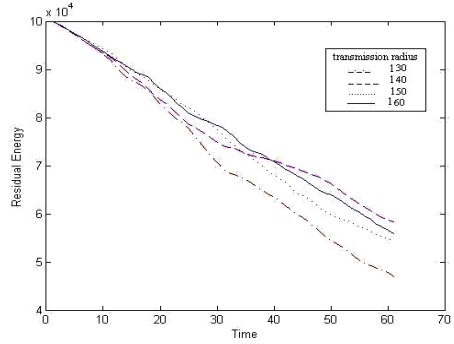

*Figure 8. Residual energy with respect different transmission radius ($r_{sense}$ = 30, Random)*

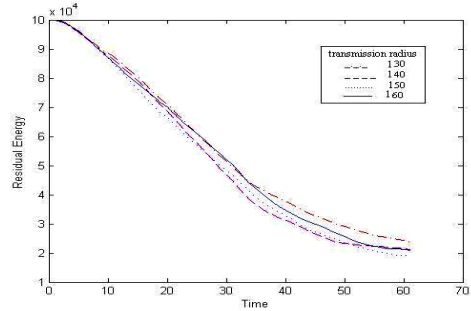

*Figure 7. Residual energy with respect different transmission radius ($r_{sense}$=30, Selective)*

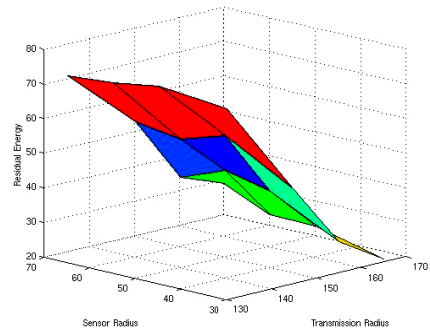

*Figure 9. Optimum sensor radius and transmission radius for Random routing method*

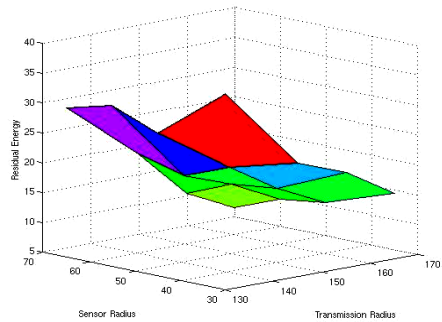

*Figure 10. Optimum sensor radius and transmission radius for Selective routing method*